\documentclass{article}
\usepackage[T1]{fontenc}
\usepackage{graphics}



\begin{document}

\title{Generic Quantum Block Compression}

\author{John Langford \thanks{This work was also partially completed
at ATnT Shannon Labs.  Address: Computer Science Department, CMU, 5000
Forbes Avenue, Pittsburgh, PA 15213}}

\maketitle
\begin{abstract}
A generic approach for compiling any classical block compression algorithm into
a quantum block compression algorithm is presented. Using this technique, compression
asymptoticaly approaching the von Neumann entropy of a qubit source can be achieved.
The automatically compiled algorithms are competitive (in time and space complexity)
with hand constructed quantum block compression algorithms.
\end{abstract}

\section{Context}

We are concerned with quantum compression, which is best understood in the context
of quantum information theory. Informally, the Schumacher coding theorem \cite{Schumacher}
states that a qubit source can be compressed to its von Neumann entropy with
high fidelity. This statement is strongly analogous to the Shannon coding theorem
for classical information theory which states that a bit source can be compressed
to its Shannon entropy with high probability. Quantum compression is therefore
concerned with algorithms for achieving compression near to the von Neumann
entropy of a qubit source in a reasonable amount of time and space.

The main result of this paper is a compilation algorithm which takes as input
any family of classical block compression (and decompression) algorithm as well
as a description of the quantum source. The output is a quantum block compression
and decompresion algorithm with the following properties:

\begin{enumerate}
\item If the classical compression algorithm asymptotically approaches Shannon entropy,
then the quantum compression algorithm asymptotically approaches the von Neumann
entropy.
\item If the classical compression and decompresion algorithm each take space \( S \)
and time \( T \), then the quantum compression algorithm uses space \( O(S\log T) \)
and time \( O(\frac{1}{\epsilon }ST^{1+\epsilon }) \) (for any \( \epsilon  \)).
\end{enumerate}
Several hand constructed algorithms for block quantum data compression have
been proposed. These include quantum arithmetic compression \cite{Ike} which
runs in \( O(n^{3}) \) time and a combinatoric index compression algorithm
\cite{Cleve} running in \( O(n^{3}) \) time and \( O(\sqrt{n}) \) space.
In further work, the combinatoric index compression was parallelized to a circuit
of only polylogarithmic depth \cite{DG}.

The remainder of this paper will discuss some information theory preliminaries,
present the compiler, compare the performance of this compiler with hand crafted
approaches, and then close with discussion.

\section{(Quantum) Information Theory}

Suppose that we have a qubit source producing qubits in the state \( |a_{i}\rangle  \)
with probability \( p_{i} \). Then the density matrix of this source will be
given by the average outer product: 
\[
\rho =\sum _{i}p_{i}|a_{i}\rangle \langle a_{i}|=\sum _{i}q_{i}|\phi _{i}\rangle \langle \phi _{i}|\]
 where \( |\phi _{i}\rangle  \) are the eigenvectors of the density matrix
and \( q_{i} \) is the eigenvalue of \( |\phi _{i}\rangle  \). 

We wish to compress the tensor product of \( n \) qubits each drawn independently
from this source so as to achieve a high ``fidelity''. Fidelity is defined
as the probability that a measurement can distinguish between the input and
output states. Let \( |b_{r}\rangle  \) be the input state chosen with random
bits \( r \), and \( |b_{r}^{'}\rangle  \) be the output of a compression/decompression
cycle. Then, the fidelity is defined by:
\[
F=E_{r}\langle b_{r}|b_{r}^{'}\rangle \]
 where \( E_{r} \) is an expectation with respect to the quantum source as
well as with respect to the outcome of the quantum measurement. 

The Schumacher coding theorem states that the minimum coding rate of any compression
scheme with a high expected fidelity is given by the von Neumann entropy:
\[
S(\rho )=Tr(\rho \log \rho )\]
Thus, we are particularly motivated by compression schemes which can (asymptotically)
achieve the von Neumann entropy.

\section{Generic quantum block compression}

A generic (classical) block compression algorithm, \( C:\{0,1\}^{n}\rightarrow \{0,1\}^{m} \),
maps \( n \) bits to \( m \) bits with \( n \) larger than \( m \). The
corresponding decompression algorithm, \( D:\{0,1\}^{m}\rightarrow \{0,1\}^{n} \),
will then (hopefully) invert the compression to give the original state. We
can use the general result that any classical function can be implemented \cite{BW}
(reasonably efficiently) on a quantum computer. In particular, If we let \( I_{1},...,I_{n} \)
be the bits of the input string and \( C_{1},...,C_{m} \) be the bits of the
compressed string then we can compile the compression circuit into the reversible
circuit which calculates: 
\[
C_{\textrm{rev}}(I_{1},...,I_{n},A_{1},...,A_{m})=I_{1},...,I_{n},A_{1}\oplus C_{1},...,A_{m}\oplus C_{m}\]
where \( A_{1},...,A_{m} \) are auxiliary bits. The same trick can be done
with the decompression circuit to get:
\[
D_{\textrm{rev}}(A_{1},...,A_{n},C_{1},...,C_{m})=A_{1}\oplus I_{1},...,A_{n}\oplus I_{n},C_{1},...,C_{m}\]
Pictorially, our circuits are the following:

\resizebox*{0.45\columnwidth}{!}{\includegraphics{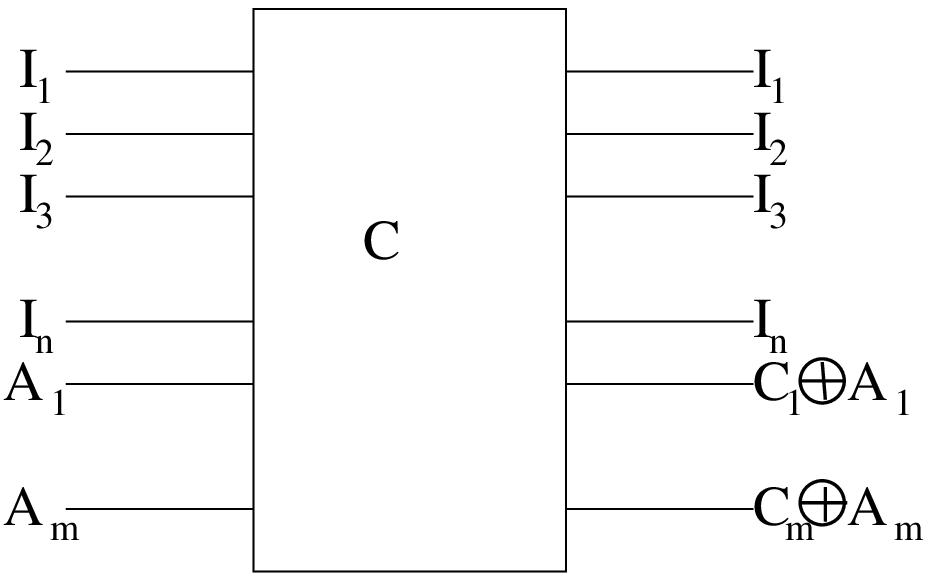}} \resizebox*{0.45\columnwidth}{!}{\includegraphics{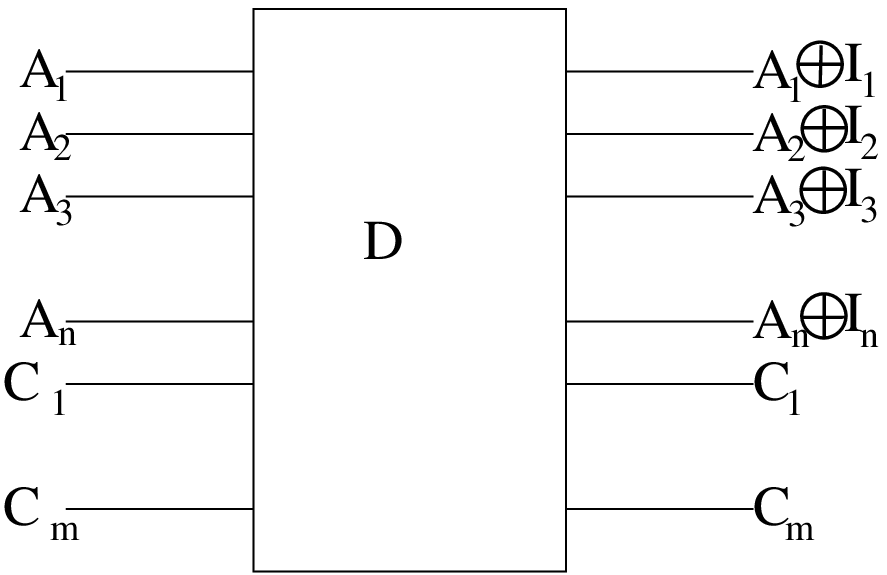}} 
\vspace{0.3cm}

These elements will be our two fundamental building blocks in the construction
of generic quantum block compression. Two challenges exist: 

\begin{enumerate}
\item \( C_{\textrm{rev}} \) and \( D_{\textrm{rev}} \) both have outputs which
are entangled with their input. This is unacceptable in any quantum compression
scheme.
\item There are some quantum sources with a low von Neumann entropy but a large Shannon
entropy. We will need to make a circuit which achieves compression near the
Shannon entropy to achieve compression near the von Neumann entropy.
\end{enumerate}
These two challenges are met in the next subsections.

\subsection{Entanglement Removal \label{unentangle}}

We are not finished with the compression because the output of \( C_{\textrm{rev}} \)
leaves the compressed bits entangled with the input bits. There is a simple
construction which removes this entanglement: compose the decompressor with
the compressor to get the following circuit:

\vspace{0.3cm}
{\par\centering \resizebox*{0.9\columnwidth}{!}{\includegraphics{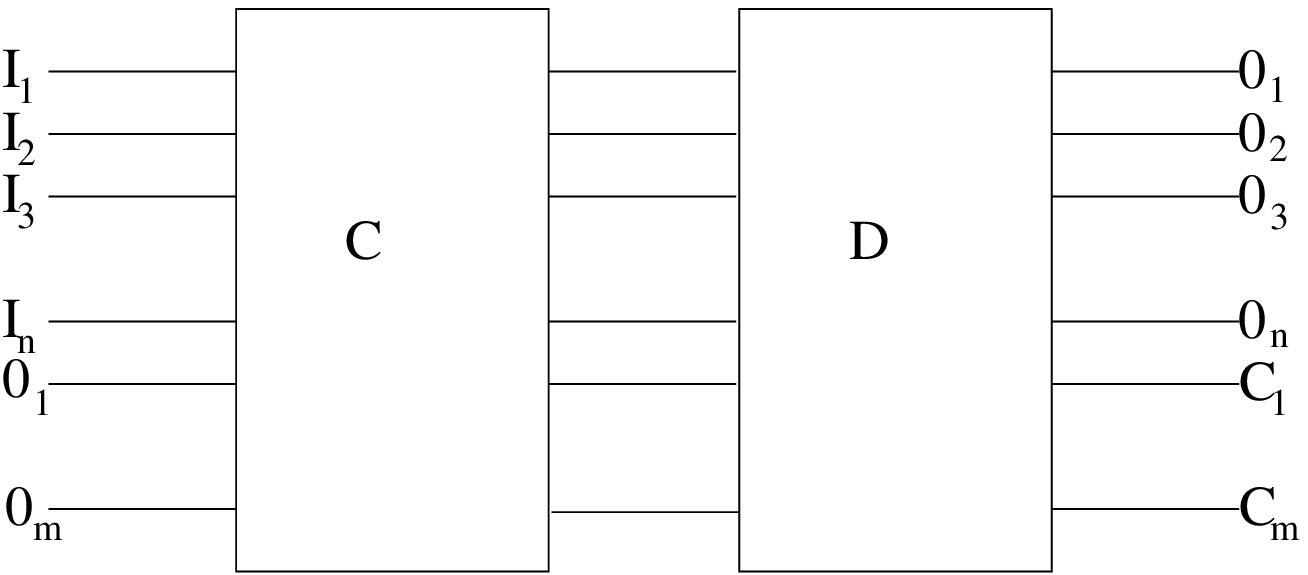}}  \par}

Here we have set the auxiliary bits to \( 0 \). 

The output of the first compression block is the compressed bits along with
the uncompressed bits. The uncompressed bits are then fed into the decompression
block instead of \( 0 \)'s, resulting in an output consisting of the compressed
bits plus things of the form \( I_{i}\oplus D(C(I))_{i} \). This term is \( 0 \)
precisely when the compression is lossless. By assumption, the compression is
lossless with high probability implying that we will essentially always measure
the \( 0 \) vector. At this point, we can measure all of the ``\( 0 \)''
qubits. If any of our measurements are nonzero, we have detected a low probability
event. The process of this measurement will project the wavefunction onto the
typical subspace, letting us achieve the projection done explicitly in \cite{Ike}
for ``free'' as a byproduct of our compression process. 

This construction is also particularly elegant because decompression (with entaglement
removal) can be accomplished by simply reading the circuit backwards. 

In summary, we now have a technique for constructing a quantum compression circuit
which outputs the compressed string \emph{without} extra entanglement.

\subsection{Achieving von Neumann entropy \label{rotate}}

The above approach only does \emph{classical} compression of qubits and so it
can only achieve the Shannon entropy. In particular, it will fail to compress
some highly compressible quantum state distributions such as the state which
is \( (|0\rangle +|1\rangle )^{n} \) with probability \( 1 \). This state
is highly compressible because it has a trivial distribution over possibilities
and yet no classical algorithm will compress the qubits well. In order to accomplish
compression of this state, we must add an extra rotation to the input and output
of the compression and decompression. For example, the operation \( H^{n} \)
(Hadamard applied to each bit), would convert the input to \( |0^{n}\rangle  \)
which is highly compressible. This extra rotation involves \( O(n) \) operations
since it can be done on each symbol from the quantum source individually giving
us the following circuit:

\vspace{0.5001cm}
{\par\centering \resizebox*{0.9\columnwidth}{!}{\includegraphics{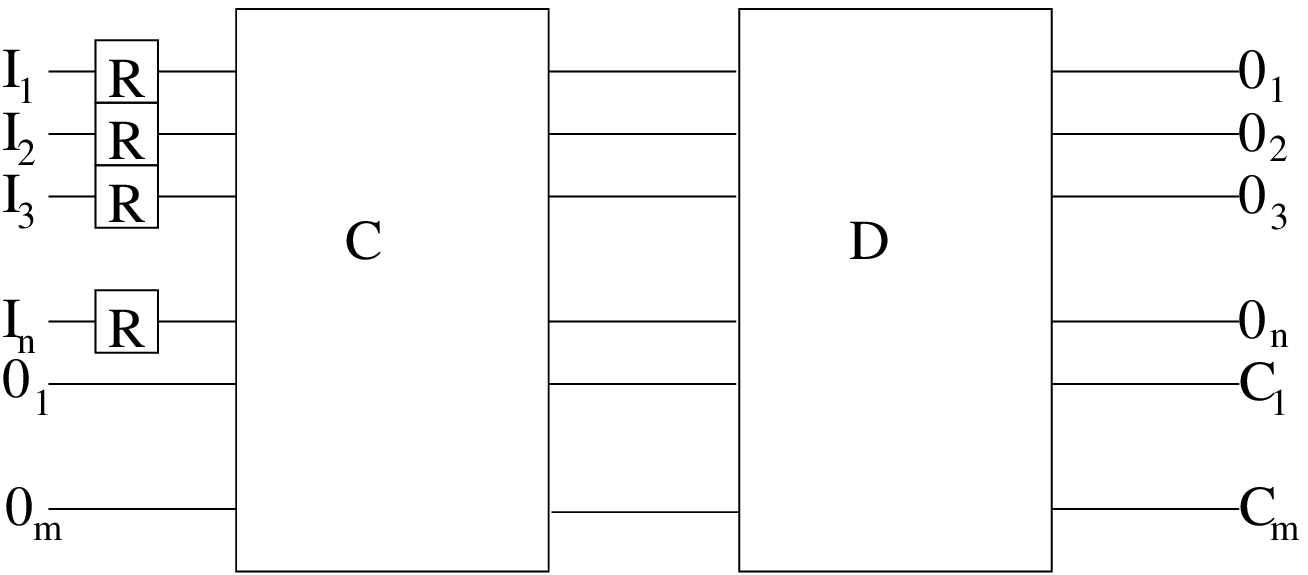}} \par}
\vspace{0.5001cm}

The interpretation of this circuit is simple: we first rotate every qubit to
align the eigenbasis and the computational basis. Then, we apply the compression
circuit using the decompression circuit to remove entanglement with the input
values. Mathematically, we compute 
\[
D_{\textrm{rev}}(C_{\textrm{rev}}(R(I_{1}),...,R(I_{n}),0_{1},..,0_{m}))\]
where \( R \) is the rotation into the eigenbasis. Once again, decompression
is done using the circuit backwards. 

This solution works for \emph{any} quantum source. Given a description of any
quantum source, the eigenvalue decomposition theorem says that we can construct
a matrix \( R \) which satisfies: 
\[
\rho =R\Lambda R^{\top }\]
 where \( \Lambda  \) is a diagonal matrix of eigenvalues and \( R \) is a
matrix consisting of eigenvectors. The \( R \) matrix will rotate the input
source so that the computational basis and the eigenbasis coincide. This rotation
does not affect the von Neumann entropy of the source. Now, given that the computational
basis and eigenbasis coincide, the von Neumann entropy will be exactly the same
as the Shannon entropy. The compiled circuit will therefore achieve the von
Neumann limit. To summarize, the compiler is the following:

\vspace{8.0pt}

\emph{Make\_quantum\_compression\_circuit(\( \rho  \), \( C \), \( D \))}

\begin{enumerate}
\item Compile the quantum reversible circuits \( D_{\textrm{rev}} \) and \( C_{\textrm{rev}} \)
from the classical circuits \( D \) and \( C \).
\item Compose \( D_{\textrm{rev}} \) and \( C_{\textrm{rev}} \) as in section \ref{unentangle}.
\item Calculate the eigenvalue matrix, \( R \), given \( \rho  \).
\item Rotate all input and output qubits of the quantum circuit by \( R \) as in
section \ref{rotate}.
\end{enumerate}

\subsection{Computational efficiency}

How efficiently can we compile compression algorithms made for classical (irreversible)
computers into reversible quantum mechanical programs? There are a very few
results here which are well summarized in \cite{TS}. There are a variety of
tradeoffs all of which are typically parameterized by the space \( S \) and
time \( T \) of the original irreversible algorithm. One of the more concrete
results is that reversible space is \( S_{\textrm{rev}}=S\log T \) and the
reversible time is (for any \( \epsilon  \)) \( T_{\textrm{rev}}=\frac{1}{\epsilon }ST^{1+\epsilon } \). 

Applying this result to the common 'bzip2' block compression algorithm which
runs in space \( S=n \) and time \( T=n\log n \) will require \( S_{\textrm{rev}}\simeq n\log n \)
space and \( T_{\textrm{rev}}\simeq \frac{1}{\epsilon }n(n\log n)^{1+\epsilon } \)
time (for any \( \epsilon  \)). If we choose \( \epsilon <1 \), This is a
smaller running time than \( O(n^{3}) \) as used by previous hand crafted quantum
block compression algorithms. It is easy to imagine that hand compilation of
irreversible compression/decompression algorithms can yield further functional
improvements.

\section{Conclusion}

The technique for automatically compiling classical compression algorithms into
quantum compression algorithms gives us a baseline and general structure for
comparison with hand crafted quantum compression algorithms. The automatic approach
is reasonably efficient and can give results better than current hand crafted
approaches in some cases. Hand crafted algorithms which are functionally more
efficient than this baseline are a nontrivial improvement. 

The technique of section \ref{unentangle} applies to \emph{any} pair of classical
algorithms with an inversion property: \( A_{2}(A_{1}(x))=x \). This includes
encryption and decryption or encoding and decoding. For algorithms with this
property, we can construct quantum analogs with a special property: the ``inputs''
are disentangled from the ``outputs''. This disentanglement comes at a relatively
small price: a factor of \( 2 \) increase in the circuit size. 

In addition to hand crafting efficient compression algorithms, one other significant
problem in quantum data compression exists: removing the requirement for full
information about the quantum source. We know this is (information theoretically)
possible \cite{Universal} although no explicit quantum algorithm yet exists
that achieves asymptotically efficient compression without knowledge of the
source. Intuitively, this is possible because the precision of the specification
of \( R \) improves exponentially with increasing (qu)bits while the necessary
precision required in specifying \( R \) scales linearly with \( n \), the
number of input qubits. Consequently, the portion of the bits which must be
allocated to specifying \( R \) is asymptotically \( 0 \). An explicit algorithm
for doing universal quantum compression would yield a significant improvement
in our understanding of quantum compression.

\section*{Acknowledgements}

I would like to thank Andris Ambainis, Klejda Bega, David Relyea, Peter Shor,
Luis von Ahn, and Ke Yang for advice and comments in preparing this paper.

\end{document}